\documentclass[12pt]{article}
\usepackage{epsf,amsfonts,amsmath}
\usepackage{graphicx,float,subfigure,epsfig,color,rotating,latexsym}
\usepackage[latin1]{inputenc}
\usepackage[T1]{fontenc}
\def\be{\begin{equation}}
\def\ee{\end{equation}} 
\def\bea{\begin{eqnarray}}
\def\eea{\end{eqnarray}} 
\def\ba{\begin{array}} 
\def\ea{\end{array}}

\def\nin{\noindent}

\begin{document}

\begin{center} 
{\bf \large{ Fluctuation-dissipation theorem and harmonic oscillators}}

\vspace*{0.8 cm}

Vincenzo Branchina\footnote{vincenzo.branchina@ct.infn.it}\label{one}
\vspace*{0.4 cm}

Department of Physics, University of
Catania and \\ INFN, Sezione di Catania, 
Via Santa Sofia 64, I-95123, Catania, Italy 

\vspace*{0.6 cm}

Marco Di Liberto\footnote{madiliberto@ssc.unict.it}\label{two}

\vspace*{0.4 cm}

Scuola Superiore di Catania, Via S. Nullo 5/i, Catania, Italy  

\vspace*{0.6 cm}

Ivano Lodato\footnote{ivlodato@ssc.unict.it}\label{three}

\vspace*{0.4 cm}
Scuola Superiore di Catania, Via S. Nullo 5/i, Catania, Italy and\\
INFN, Sezione di Catania, Via Santa Sofia 64, I-95123, Catania, Italy 

\vspace*{1.2 cm}

{\LARGE Abstract}\\

\end{center}
The question of the ``physical meaning'' and ``origin'' of the 
Bose-Einstein (BE) factor in the fluctuation-dissipation theorem 
(FDT) is often raised and this term is sometimes 
interpreted as originating from a real harmonic oscillator 
composition of the physical system. Such an interpretation, 
however, is not really founded. Inspired by the famous 
work of Caldeira and Leggett, we have been able to show that, 
whenever linear response theory is applicable, which is the 
main hypothesis under which the FDT is established, any generic 
bosonic and/or fermionic system at temperature $T$ can be mapped 
onto a fictitious system of harmonic oscillators so that the 
suscettivity and the mean square of the fluctuating observable 
of the real system coincide with the corresponding quantities  
of the fictitious one. We claim that it is in this sense, and only 
in this sense, that the BE factor can be interpreted in terms of 
harmonic oscillators, no other physical meaning can be superimposed 
to it. At the best of our knowledge, this is the first time that 
such a mapping is explicitly worked out.

\vspace*{1.6cm}

The fluctuation-dissipation theorem (FDT)\,\cite{cawe} is very general 
and applies to a broad variety of different physical phenomena. 
An often raised question concerns the physical meaning and/or origin 
of the Bose-Einstein (BE) distribution factor which appears in the 
relation between the power spectrum of the fluctuating quantity and 
the corresponding generalized suscettivity. 
Sometimes this term is interpreted as due to an harmonic oscillator 
composition of the considered physical system. Such an interpretation, 
however, is not supported by the derivation of the theorem itself. 
Moreover, the FDT applies to any generic bosonic and/or fermionic
system. 

Far from being an academic question, this issue is of very practical 
importance in many different contexts ranging from solid state 
physics to astrophysics and cosmology. Actually, from a real 
understanding of the origin of this term often depends the correct 
physical interpretation of theoretical and experimental results 
(see below). In this letter we solve this important interpretation 
issue.

To this end, inspired by seminal papers of Caldeira 
and Leggett\,\cite{caleg1, caleg2}, in the following we shall be able 
to establish a new very general result (a mapping between any generic
system at temperature $T$ and a fictitious system of harmonic oscillators), 
whose interest goes beyond the specific application to the FDT 
investigated in the present work.  

Given a generic bosonic or fermionic system which interacts with an 
external field $f(t)$ through the interaction term 
$\hat V = - f(t)\,\hat A$, where 
$\hat A$ is an observable (a bosonic operator) of the system, the 
fluctuation-dissipation theorem states that, whenever linear response 
theory is applicable, the mean square of the Fourier transform 
$\hat A(\omega)$ of $\hat A(t)$ is related to the imaginary 
part $\chi_{_A}''(\omega)$ of the corresponding (Fourier transformed) 
generalized suscettivity by the relation\,\cite{cawe} :
\be \label{fdt1} 
\langle \hat{A}^2(\omega)\rangle =
\hbar \chi_{_A}^{\,''}(\omega) \,{\rm coth}\left( \frac{\beta\hbar\omega}{2}
\right) =2\,\hbar\, \chi_{_A}^{\,''}(\omega) \,\left(\frac1 2 + \frac{1}
{e^{\beta\hbar\omega}-1}\right)\,,
\ee
where $\beta=1/k T$, with $T$ the temperature of the system and $k$ the 
Boltzmann constant.

Let us consider, for instance, a resistively shunted Josephson 
junction\,\cite{koch} and, more specifically, the power spectrum 
$S_I(\omega)$ of the noise current (fluctuation) in the resistive 
shunt (dissipation). 
When applied to this case, the theorem takes the form 
($R$ is the shunt resistance)\cite{koch1}:
\be
\label{spectr}
S_I(\omega) = \frac{4}{R}\left(\frac{\hbar\omega}{2}+
\frac{\hbar\omega}{e^{\beta\hbar\omega}-1}\right)\, .
\ee

The power spectrum $S_I(\omega)$ has been measured\,\cite{koch}
and good agreement between the experimental results and 
Eq.\,(\ref{spectr}) has been found. 
The specialized literature\,\cite{koch, koch1, gardizo, kogan} often 
presents the $\frac{\hbar\omega}{2}$ 
term in parenthesis as due to zero point energies and the  
experimental results\,\cite{koch} as a measurement of
zero point energies. In fact, the term in parenthesis in 
Eq.\,(\ref{spectr}) coincides with the mean energy 
of an harmonic oscillator of frequency $\omega$ in a thermal bath. The 
same holds true for the general case of Eq.\,(\ref{fdt1}),
where the term in parenthesis is the mean energy of an harmonic 
oscillator in $\hbar\omega$ units, i.e.\,the BE
distribution function. 

The question that we would like to answer is if it is physically 
founded to interpret this BE term as coming from 
an harmonic oscillator structure of the system. In particular, we 
would like to understand if the agreement between the experimental 
results\,\cite{koch} and Eq.\,(\ref{spectr}) can be considered as a 
signature of measurement of zero point energies. 

We anticipate now the results of our analysis. 
On the one hand, we shall be able to establish a {\it mapping} 
between the physical system and a fictitious system of harmonic 
oscillators in such a manner that the mean square of 
$\hat A(\omega)$ and the related (imaginary part of the) 
generalized suscettivity $\chi_{_A}^{\,''}(\omega)$ of the 
(real) system are precisely reproduced by considering the 
corresponding quantities of the fictitious one. 
At the same time, our analysis will clarify that it is only in 
this sense that the BE factor can be interpreted in terms of 
harmonic oscillators. Therefore, no physical oscillator degrees 
of freedom of the system are involved in Eq.\,(\ref{fdt1}) and 
no zero point energies have been measured in\,\cite{koch}. 

Let us begin by briefly reviewing the derivation 
of the FDT. Consider a macroscopic system with unperturbed 
Hamiltonian $\hat{H}_0$ under the influence of the perturbation 
\be
\label{inter}
\hat{V} = - f(t)\,\hat {A}(t)\,,
\ee
where $\hat{A}(t)$ is an observable (a bosonic operator) of 
the system and $f(t)$ an external 
generalized force\footnote{More generally, we could consider 
a local observable and a local generalized force, in which case 
we would have $\hat{V} = -\int d^3\,\vec r \hat{A}(\vec{r})f(\vec{r},t)$, 
and successively define a local suscettivity 
$\chi(\vec{r},t;\vec{r'},t')$ (see Eq.\,(\ref{chi}) below).  
As this would add nothing to our argument, we shall restrict ourselves
to $\vec r$-independent quantities. The extension 
to include local operators is immediate.}. 
Let $|E_n\rangle$ be the $\hat{H}_0$ 
eigenstates (with eigenvalues $E_n$) and  
$\langle E_n|\hat{A}(t)|E_n \rangle =0$. 
Within the framework of linear response theory, the quantum-statistical 
average $\langle\hat{A}(t)\rangle_f$ of the observable $\hat{A}(t)$ 
in the presence of $\hat{V}$ is given by 
\be
\label{resp2}
\langle \hat{A}(t)\rangle_f = \int_{-\infty}^t\, d t' \chi_{_{A}}(t-t') f(t') 
\ee
where $\chi_{_{A}}(t - t')$ is the generalized suscettivity,
\be
\label{chi}
\chi_{_{A}}(t - t')=\frac{i}{\hbar}\theta(t-t') 
\langle [\hat{A}(t),\hat{A}(t')] \rangle = 
-\frac{1}{\hbar}G_R(t - t')\, ,
\ee
with $\langle ... \rangle = 
\sum_{n} \varrho_n \langle E_n| ... |E_n \rangle$,\, 
$\varrho_n= e^{-\beta E_n}/Z$\, ,  $Z=\sum_n e^{- \beta E_n}$\,, 
$G_R(t-t')$ being the retarded Green's function and 
$\hat{A}(t)=e^{i\hat{H_0}t/\hbar}\hat{A}e^{-i\hat{H_0}t/\hbar}$.

Defining the correlators (from now on  $t^{'}=0$):
\be\label{correla}
G_{>}(t)= \langle \hat{A}(t)\,\hat{A}(0) \rangle 
\,\,\,\,\,\,\,\,\,\,\,\,\,\,\, {\rm and} \,\,\,\,\,\,\,\,\,\,\,\,\,\,\, 
G_{<}(t)= \langle \hat{A}(0)\,\hat{A}(t) \rangle\,, 
\ee
so that $ G_R(t)=-i\theta(t)(G_{>}(t) - G_{<}(t))$, 
and the corresponding Fourier transforms,
$ G_{>}(\omega)$ and $ G_{<}(\omega)$ respectively, it is a matter of 
few lines to show that:
\be\label{aux}
 G_{>}(\omega)=-\frac{2}{1-e^{-\beta\hbar\omega}}{\rm Im}\,G_R(\omega)
\,\,\,\,\,\,\,\,\,\,\,\,\,\,\, ; \,\,\,\,\,\,\,\,\,\,\,\,\,\,\, 
G_{<}(\omega)=e^{-\beta\hbar\omega}\,G_{>}(\omega) \,.
\ee
Finally, by noting that  
\be\label {oo2}
\langle \hat{A}^2(\omega)\rangle= \frac12 (G_{>}(\omega) + G_{<}(\omega)) 
\ee
and that the Fourier transform of $\chi_{_{A}}(t)$\, is\, 
$\chi_{_{A}}(\omega) = \chi_{_{A}}^{\,'}(\omega) + i \chi_{_{A}}^{\,''}(\omega) = 
-\frac{1}{\hbar} G_{R}(\omega)$ we get:
\be \label{fddt} 
\langle \hat{A}^2(\omega)\rangle = \hbar \chi_{_{A}}^{\,''}(\omega) \, \frac
{1 + e^{- \beta\hbar\omega}}
{1 - e^{- \beta\hbar\omega}} =
\hbar \chi_{_{A}}^{\,''}(\omega) \,{\rm coth}\left( \frac{\beta\hbar\omega}{2}\right)
=2\,\hbar \chi_{_{A}}^{\,''}(\omega) \,
\left(\frac1 2 + \frac{1}{e^{\beta\hbar\omega}-1}\right) 
\ee
which is Eq.\,(\ref{fdt1}), the celebrated FDT. 

As observed by Kubo et al.\,\cite{kubo} (and shown in the derivation sketched 
above), the BE factor is simply due to a peculiar combination of Boltzmann 
factors in Eq.\,(\ref{fddt}) and there is no reference to physical harmonic 
oscillators of the system whatsoever.
Despite such an authoritative remark, some people insist in interpreting 
the $\left(\frac1 2 + \frac{1}{e^{\beta\hbar\omega}-1}\right)$ term as related 
to an harmonic oscillator structure of the physical system\footnote{In the 
case of the measured\,\cite{koch} power spectrum of Eq.\,(\ref{spectr}), 
some authors\,\cite{bema1, bema2} interpret this term as due to the 
electromagnetic field in the resistive shunt and therefore the first term 
in parenthesis of Eq.\,(\ref{spectr}) as originating from zero point energies 
of this electromagnetic field.}. 

To begin with our analysis, it is useful to show that from Eqs.\,(\ref{chi}) 
and (\ref{oo2}) we can easily derive the following expressions for 
$\chi_{_A}^{\,''}(\omega)$ and $\langle \hat{A}^2(\omega)\rangle$ :      
\be\label{chi2}
\chi_{_A}''(\omega)=\frac{\pi}{\hbar}\sum_{i, j}\varrho_i |A_{i j}|^2
\left[\delta\left(\frac{E_i-E_j}{\hbar}+\omega\right)-
\delta\left(\frac{E_j-E_i}{\hbar}+\omega\right)\right]\,,
\ee
and 
\be\label{o2}
\langle \hat{A}^2(\omega)\rangle=\pi\sum_{i, j}\varrho_i |A_{ij}|^2
\left[\delta\left(\frac{E_i-E_j}{\hbar}+\omega\right)+
\delta\left(\frac{E_j-E_i}{\hbar}+\omega\right)\right]\,,
\ee
where $A_{ij}=\langle E_i|\hat A | E_j\rangle $ . 

In fact, by inserting in Eq.\,(\ref{chi}) the expressions:\hfill\break
$\theta(t-t')=-\int_{-\infty}^{+\infty}\frac{d\omega}
{2\pi i}\frac{e^{-i\omega(t-t')}}{\omega+i\eta}$ , $I=\sum_i |E_i\rangle\langle E_i|$
and $\hat{A}(t)=e^{i\hat{H_0}t/\hbar}\hat{A}e^{-i\hat{H_0}t/\hbar}$ we get:
\begin{equation}
\chi_{_A}(t-t')=-\frac1 \hbar \int \frac{d\omega}{2\pi}\frac{e^{-i\omega(t-t')}}{\omega+i\eta}
\sum_{i, j} \varrho_i |A_{ij}|^2
\left(e^{i(E_i-E_j)(t-t')/\hbar}- e^{-i(E_i-E_j)(t-t')/\hbar} \right)
\end{equation}
Then, by defining the Bohr frequencies $\omega_{ij}=(E_i-E_j)/\hbar$ and making 
use of the relation 
$\lim_{\eta\rightarrow 0}\frac{1}{\omega+i\eta}= \mathcal{P} 
\left(\frac1 \omega\right)  - i\pi\delta(\omega)$
we have for $\chi_{_A}^{''}(t-t')$:

\bea
\chi_{_A}^{''}(t-t')&=&\frac{\pi}{\hbar}\int\frac{d\omega}
{2\pi}\sum_{i, j} \varrho_i |A_{ij}|^2\left( e^{-i(\omega-\omega_{ij})(t-t')}\delta(\omega) 
- e^{-i(\omega+\omega_{ij})(t-t')}\delta(\omega) \right) \nonumber\\
&=&\frac{\pi}{\hbar}\int\frac{d\omega}{2\pi}e^{-i\omega (t-t')}
\sum_{i, j} \varrho_i|A_{ij}|^2 \left(\delta(\omega+\omega_{ij})
-\delta(\omega + \omega_{ji})\right)
\eea
which immediately gives Eq.\,(\ref{chi2}).

As for Eq.\,(\ref{o2}), by inserting the identity 
in Eq.\,(\ref{correla}) for $G_>(t)$ we find: 
\begin{eqnarray}\label{uno}
G_>(t)&=&\langle \hat{A}(t)\hat{A}(0)\rangle= \sum_{i,j}\rho_i
\langle E_i| e^{\frac i \hbar \hat H t}\hat A 
e^{-\frac i \hbar \hat H t}|E_j\rangle\langle E_j|\hat A|E_i\rangle=\nonumber \\ 
&=&\sum_{i,j}\rho_i e^{-\frac i \hbar(E_j - E_i)t}|A_{ij}|^2\,.
\end{eqnarray}
Working out the similar expression for $G_<(t)$, for the correlation function 
$G(t)$ we finally have: 

\begin{equation}\label{due}
G(t)=\frac1 2 (G_>(t) + G_<(t))=\frac1 2 \sum_{i , j}\rho_i |A_{ij}|^2 
(e^{-\frac i \hbar(E_j - E_i)t}+e^{-\frac i \hbar(E_i - E_j)t})\, ,
\end{equation}
and taking the Fourier transform: 
\be\label{ft}
\tilde G(\omega) =\pi\sum_{i , j}\rho_i|A_{ij}|^2\left[\delta\left(\frac{E_j-E_i}{\hbar}+
\omega\right)+\delta\left(\frac{E_i-E_j}{\hbar} +\omega\right) \right]\,.
\ee
As 
\be\label{giti}
G(t)=\int_{-\infty}^{+\infty} \tilde G(\omega) e^{-i\omega t}\frac{d\omega}{2\pi}
\ee 
and 
$\langle\hat{A}^2\rangle = G(0)$, $\tilde G(\omega)$ is nothing but the spectral 
density $\langle \hat{A}^2(\omega)\rangle$ of $\langle\hat{A}^2\rangle$ : 
\be\label{specde}
\langle \hat{A}^2\rangle=\int_{-\infty}^{+\infty}\langle \hat{A}^2(\omega)
\rangle\frac{d\omega}{2\pi}\,.
\ee
From Eqs.\,(\ref{ft}), (\ref{giti}) and (\ref{specde}), we 
immediately get Eq.\,(\ref{o2}). 

\nin
For our purposes, it is useful to write Eq.\,(\ref{o2}) in
a different manner. It is not difficult to show that we can actually 
write Eq.\,(\ref{o2}) as\footnote{Obviously, comparing Eq.\,(\ref{o4}) with 
Eq.\,(\ref{chi2}), we find, as we should, the FDT theorem, but this is not 
our goal.}: 
\bea
\langle \hat{A}^2(\omega)\rangle &=& \pi\sum_{i, j}\varrho_i |A_{ij}|^2
\,{\rm coth}\left( \frac{\beta\hbar\omega_{ji}}{2}\right)
\left[\delta\left(\omega_{ij}+\omega\right)+
\delta\left(\omega_{ji}+\omega\right)\right]\nonumber\\
&=& \pi\,{\rm coth}\left( \frac{\beta\hbar\omega}{2}\right)\sum_{i, j}\varrho_i |A_{ij}|^2
\left[\delta\left(\omega_{ij}+\omega\right)-
\delta\left(\omega_{ji}+\omega\right)\right]\,.\label{o4}
\eea
After few additional steps we finally have: 
\bea
\langle \hat{A}^2(\omega)\rangle &=& \pi\sum_{j > i}(\varrho_i - \varrho_j) |A_{ij}|^2
\,{\rm coth}\left( \frac{\beta\hbar\omega_{ji}}{2}\right)
\left[\delta\left(\omega -\omega_{ji}\right)+
\delta\left(\omega + \omega_{ji}\right)\right]\label{o5}\\
&=& \pi\,{\rm coth}\left(\frac{\beta\hbar\omega}{2}\right)\sum_{j > i}
(\varrho_i - \varrho_j) |A_{ij}|^2 \left[\delta\left(\omega -\omega_{ji}\right)-
\delta\left(\omega + \omega_{ji}\right)\right]\,.\label{o6}
\eea
Similarly, it is easy to see that Eq.\,(\ref{chi2}) can be written as:  
\be\label{chii3}
\chi''(\omega)=\frac{\pi}{\hbar}\sum_{j > i}(\varrho_i - \varrho_j) |A_{i j}|^2
\left[\delta\left(\omega - \omega_{ji}\right)-
\delta\left(\omega + \omega_{ji}\right)\right]\,. 
\ee

Starting from Eqs.\,(\ref{o5}) and (\ref{chii3}) and taking inspiration 
from previous work of Caldeira and Leggett\,\cite{caleg2}, we shall be able 
to establish a formal mapping between the real system considered so far 
and a system of fictitious harmonic oscillators. 
A similar mapping, restricted however to the $T=0$ temperature case, 
was considered in\,\,\cite{caleg2}, where it was also noted that 
the $T\neq 0$ case needs separate discussion. 
The mapping that we are going construct in the present work actually 
deals with the $T\neq 0$ general case. At the best of our knowledge, 
this is the first time that such a mapping is explicitly worked out.

To prepare the basis for the construction of this mapping, let us 
consider first a real system ${\cal S}_{osc}$ of 
harmonic oscillators (each of which is labeled below by the double 
index $\{ji\}$ for reasons that will become clear in the following) 
whose free Hamiltonian is: 
\be\label{armonico}
\hat H_{osc} = \sum_{j > i}\left(\frac{\hat p_{ji}^{\,2} }{2 M_{ji}} + 
\frac{M_{ji} \omega_{ji}^2}{2}\,\hat q_{ji}^{\,2}\right)\,,
\ee
where $\omega_{ji}$ are the proper 
frequencies of the individual harmonic oscillators and $M_{ji}$ their 
masses. Let  $| n_{j i}\rangle$ ($n_{ji}=0, 1,2,...$) be the 
occupation number states of the $\{ji\}$ oscillator out of which the Fock 
space of ${\cal S}_{osc}$ is built up.
Let us consider also ${\cal S}_{osc}$ in interaction with an external 
system through the one-particle operator: 
\be\label{armint}
\hat V_{osc} = - f(t) \hat A_{osc}\,, 
\ee
with 
\be\label{onepart}
{\hat A}_{osc} = \sum_{j > i} \left(\alpha_{j i} \,{\hat q}_{ji} \right)\,.
\ee
Obviously, the FDT applied to ${\cal S}_{osc}$ gives: 
\be \label{fdtosc} 
\langle {\hat A}_{osc}^2(\omega)\rangle = 
\hbar \chi_{osc}^{\,''}(\omega) \,{\rm coth}\left( \frac{\beta\hbar\omega}{2}\right)\,,
\ee
but this is not what matters to us. 

What is important for our purposes is that, differently from 
any other generic system, for ${\cal S}_{osc}$ we can exactly compute 
$\langle {\hat A}_{osc}^2(\omega) \rangle$ and $\chi_{osc}^{\,''}(\omega)$
from Eqs.\,(\ref {chi2}) and (\ref{o2}) because we can explicitly compute 
the matrix elements of ${\hat A}_{osc}$. 

In fact, if we apply Eqs.\,(\ref{uno}), (\ref{due}) and (\ref{ft}) 
to ${\cal S}_{osc}$ and replace the double index 
notation introduced above 
($\omega_{ji}$ ; $n_{ji}$ ; $M_{ji}$ ;  etc.) with the more 
convenient (for the time being) and self explanatory one index 
notation ($\omega_{1}$, $\omega_{2}$, ... ; 
$n_{1}$, $n_{2}$, ... ; $M_{1}$, $M_{2}$, ... ; etc.), for 
$\langle \hat{A}_{osc}^2(\omega)\rangle$ we have:
\bea\label{osc}
\langle \hat{A}_{osc}^2(\omega)\rangle&=&\pi
\sum_{n_1,n_2,..}\,\,\,\,  \sum_{m_1,m_2,..}\,(\varrho_{n_1}  \varrho_{n_2}\cdots)
|\langle n_1,n_2,..|\hat{A}_{osc} | m_1,m_2,..\rangle|^2 \nonumber\\
&&\times
\left[\delta \left(\omega + l_1\omega_1 + l_2\omega_2 +\cdots \right)+ 
\delta \left(\omega - l_1\omega_1 - l_2\omega_2 - \cdots \right)\right]
\eea
where $l_k=n_k-m_k$, $\varrho_{n_k} = e^{-\beta (n_k + 1/2)\hbar\omega_k}/Z_k$, 
$Z_k = \sum_{n_k} e^{-\beta (n_k + 1/2)\hbar\omega_k}$ (note also that in this 
one index notation  $\hat{A}_{osc}$ is written as \, 
${\hat A}_{osc} = \sum_{k} \left(\alpha_{k} \,{\hat q}_{k} \right)$\,). 
Now, as
\be
\langle n_k| {\hat q}_{k} | m_k\rangle = \sqrt {\frac{\hbar}{2 M_k \omega_k}}
\left(\sqrt {n_k + 1} \langle n_k + 1| m_k \rangle + 
\sqrt {n_k} \langle n_k - 1| m_k \rangle \right) \,, 
\ee
we immediately get:
\bea\label{osc2}
\langle \hat{A}_{osc}^2(\omega)\rangle&=&\pi
\sum_{n_1,n_2,..}\,\,\,\,  \sum_{m_1,m_2,..}\,(\varrho_{n_1}  \varrho_{n_2}\cdots)
~~~~~~~~~~~~~~~~~~~~~~~~~~~~~~~~~~~~~~~~~~~~~~~~~~~~~~~~~~~~\nonumber\\
&\times&\left[\sum_k \alpha_k \sqrt {\frac{\hbar}{2 M_k \omega_k}}
\left(\sqrt {n_k + 1} \delta_{m_k, n_k+1} + \sqrt {n_k} \delta_{m_k, n_k-1}\right)
\prod_{h\neq k} \delta_{m_h, n_h}\right]^2 \nonumber\\
&\times&
\left[\delta \left(\omega + l_1\omega_1 + l_2\omega_2 +\cdots \right)+ 
\delta \left(\omega - l_1\omega_1 - l_2\omega_2 - \cdots \right) \right]\,.
\eea

Let us concentrate our attention to the square in 
the second line of Eq.\,(\ref{osc2}). Due to the presence of the 
Kronecker deltas,  
all the crossed terms in this square, i.e. all the terms with different 
values of the index $k$, vanish. In other words, the square of the sum 
is equal to the sum  of the squares:
\bea
\left[\sum_k \alpha_k \sqrt {\frac{\hbar}{2 M_k \omega_k}}
\left(\sqrt {n_k + 1} \delta_{m_k, n_k+1} + \sqrt {n_k} \delta_{m_k, n_k-1}\right)
\prod_{h\neq k} \delta_{m_h, n_h}\right]^2 \nonumber\\
= \sum_k \left(\alpha_k \sqrt {\frac{\hbar}{2 M_k \omega_k}}
\left(\sqrt {n_k + 1} \delta_{m_k, n_k+1} + \sqrt {n_k} \delta_{m_k, n_k-1}\right)
\prod_{h\neq k} \delta_{m_h, n_h}\right)^2
\eea
For the same reason, the same holds true
for each value of the index $k$, i.e.:
\bea
&&\left(\alpha_k \sqrt {\frac{\hbar}{2 M_k \omega_k}}
\left(\sqrt {n_k + 1} \delta_{m_k, n_k+1} + \sqrt {n_k} \delta_{m_k, n_k-1}\right)
\prod_{h\neq k} \delta_{m_h, n_h}\right)^2 \nonumber\\
&=&\alpha_k^2\, \frac{\hbar}{2 M_k \omega_k}
\left( (n_k + 1)\, \delta_{m_k, n_k+1} + n_k\, \delta_{m_k, n_k-1}\right)
\prod_{h\neq k} \delta_{m_h, n_h}\label{osscc}\,.
\eea
Therefore, as $l_k=n_k-m_k $, for 
$\langle \hat{A}_{osc}^2(\omega)\rangle$ we get:
\be\label{osc3}
\langle \hat{A}_{osc}^2(\omega)\rangle=\pi
\sum_{n_1,n_2,..}\, (\varrho_{n_1}  \varrho_{n_2}\cdots)\,
\sum_k \alpha_k^2\, \frac{\hbar}{2 M_k \omega_k}
(2\,n_k + 1)\, (\delta (\omega-\omega_k) +\delta (\omega+\omega_k) )
\ee
Finally, as $ \sum_{n_k}\, \varrho_{n_k}=1 $, the above expression becomes: 
\bea
\langle \hat{A}_{osc}^2(\omega)\rangle&=&\pi
\sum_k \, \alpha_k^2\, \frac{\hbar}{2 M_k \omega_k}
(\delta (\omega-\omega_k) +\delta (\omega+\omega_k))
\sum_{n_k}\, \varrho_{n_k}\,(2\,n_k + 1) \label{si}\\
&=&\pi
\sum_k \, \alpha_k^2\, \frac{\hbar}{2 M_k \omega_k}\,
{\rm coth}\left(\frac{\beta \hbar\omega_k}{2}\right)
(\delta (\omega-\omega_k) +\delta (\omega+\omega_k))\label{osc4}\,.
\eea
Going back to the original double index notation, the above equation 
is written as: 
\bea
\langle \hat{A}_{osc}^2(\omega)\rangle&=&\pi
\sum_{j>i} \, \alpha_{ji}^2\, \frac{\hbar}{2 M_{ji} \omega_{ji}}\,
{\rm coth}\left(\frac{\beta \hbar\omega_{ji}}{2}\right)
(\delta (\omega-\omega_{ji}) +\delta (\omega+\omega_{ji}))\label{osc5}\\
&=& \pi\, {\rm coth}\left(\frac{\beta \hbar\omega}{2}\right)\sum_{j>i} \,
\alpha_{ji}^2\,\frac{\hbar}{2 M_{ji} \omega_{ji}}\,
(\delta (\omega-\omega_{ji}) - \delta (\omega+\omega_{ji}))\label{oscc5}\,.
\eea

We have just seen that given a real system ${\cal S}_{osc}$ of harmonic 
oscillators and the one particle operator $\hat{A}_{osc}$  of 
Eq.\,(\ref{onepart}), for such an operator is possible to evaluate explicitly 
$\langle \hat{A}_{osc}^2(\omega)\rangle$. We find that each of the individual 
harmonic oscillators gives rise to a term  
${\rm coth}\left(\frac{\beta \hbar\omega_{ji}}{2}\right)$ which in turn 
comes from the term $\sum_{n_{ji}}\, \varrho_{n_{ji}}\,(2\,n_{ji} + 1)$
of Eq.\,(\ref{si}).

Let us now consider $\chi_{osc}^{''} (\omega)$,  which   
(see Eqs.\,(\ref{chi2}) and (\ref{osc2})) is nothing but:
\bea\label{cchi2}
\chi^{''}_{osc} (\omega)&=&\frac{\pi}{\hbar}
\sum_{n_1,n_2,..}\,\,\,\,  \sum_{m_1,m_2,..}\,(\varrho_{n_1}  \varrho_{n_2}\cdots)
~~~~~~~~~~~~~~~~~~~~~~~~~~~~~~~~~~~~~~~~~~~~~~~~~~~~~~~~~~~~\nonumber\\
&\times&\left[\sum_k \alpha_k \sqrt {\frac{\hbar}{2 M_k \omega_k}}
\left(\sqrt {n_k + 1} \delta_{m_k, n_k+1} + \sqrt {n_k} \delta_{m_k, n_k-1}\right)
\prod_{h\neq k} \delta_{m_h, n_h}\right]^2 \nonumber\\
&\times&
\left[\delta \left(\omega + l_1\omega_1 + l_2\omega_2 +\cdots \right) -  
\delta \left(\omega - l_1\omega_1 - l_2\omega_2 - \cdots \right) \right] .
\eea
Apart from the factor $1/\hbar$, Eq.\,(\ref{cchi2}) differs from Eq.\,(\ref{osc2})
because it contains the difference (rather than the sum) of delta functions 
in the last line.  

If we proceed for $\chi_{osc}^{''} (\omega)$ as we have just done for 
$\langle \hat{A}_{osc}^2(\omega)\rangle$, we immediately note that 
the only difference with the previous computation is due to this minus 
sign. In fact, its presence causes that rather than the 
combination  $(2\,n_k + 1)$ of Eq.\,(\ref{osc3}), which comes from the sum 
$(n_k + 1) + n_k$ of Eq.\,(\ref{osscc}), we get the combination
$(n_k + 1) - n_k = 1$. Therefore, for $\chi_{osc}^{''} (\omega)$ we
do not get the sum 
$\sum_{n_k}\, \varrho_{n_k}\,(2\,n_k + 1)= 
{\rm coth}\left(\frac{\beta \hbar\omega_k}{2}\right)$ of Eq.\,(\ref{si}), 
but rather\, $\sum_{n_k}\, \varrho_{n_k}\, = 1$. Then: 
\be\label{chi3}
\chi_{osc}^{''} (\omega)=\frac{\pi}{\hbar}
\sum_{j>i} \,\alpha_{ji}^2\,\frac{\hbar}{2 M_{ji} \omega_{ji}}\,
(\delta (\omega-\omega_{ji}) - \delta (\omega+\omega_{ji}))\,.
\ee

Naturally, comparing Eq.\,(\ref{oscc5}) with Eq.\,(\ref{chi3}) we 
see that for ${\cal S}_{osc}$ the FDT holds true, 
as it should. However, what is important for our purposes is to note 
that for this system we have been able to compute separately 
$\langle \hat{A}_{osc}^2(\omega)\rangle$ and $\chi_{osc}^{''} (\omega)$  
and found that the 
${\rm coth}\left(\frac{\beta \hbar\omega}{2}\right)$ factor of the FDT 
originates from the individual contributions 
${\rm coth}\left(\frac{\beta \hbar\omega_{ji}}{2}\right)$ of each of 
the harmonic oscillators of ${\cal S}_{osc}$.

We are now in the position to build up our mapping. 
Let us consider the original system ${\cal S}$, described by the 
unperturbed Hamiltonian $\hat H_0$\,, in interaction with an external 
field $f(t)$ through the interaction term  $\hat V = - f (t)\,\hat A$  
(see Eq.\,(\ref{inter})), and construct a fictitious system of harmonic 
oscillators ${\cal S}_{osc}$, described by the free Hamiltonian 
${\hat H}_{osc}$ of Eq.\,(\ref{armonico}), in interaction with the same 
external field $f(t)$ through the interaction term ${\hat V}_{osc}$ of 
Eq.\,(\ref{armint}), with $\hat A_{osc}$ given by Eq.\,(\ref{onepart}),  
where for $\alpha_{j i}$ we choose:
\be\label{alfa}
\alpha_{j i} = \left(\frac{2 M_{ji} \omega_{ji}}{\hbar}\right)^{\frac12}
(\varrho_i - \varrho_j)^{\frac12} \,|A_{ij}|
\ee
and for the proper frequencies $\omega_{ji}$ of the oscillators: 
\be\label{omega}
\omega_{ji}= (E_j-E_i)/\hbar > 0\,,
\ee
with $E_i$ the eigenvalues of the Hamiltonian ${\hat H}_0$ of the real system. 

Comparing Eq.\,(\ref{chi3}) with Eq.\,(\ref{chii3}) and Eq.\,(\ref{oscc5}) with 
Eq.\,(\ref{o6}), it is immediate to see that 
with the above choices of $\alpha_{j i}$ and $\omega_{ji}$ we have: 
\bea
\langle \hat{A}^2(\omega)\rangle &=& 
\langle \hat{A}_{osc}^2(\omega)\rangle\label{cen1}\\
\chi_{_A}^{''} (\omega) &=& \chi_{osc}^{''} (\omega)\label{cen2}\,.
\eea

Eqs.\,(\ref{alfa}) and (\ref{omega}) are the central results of our 
analysis. These are the equations that allow to establish the desired 
mapping. Actually, with such a choice of 
the $\alpha$'s and the $\omega$'s, we have been able to derive 
Eqs.\,(\ref{cen1}) and (\ref{cen2}).
What we have just found is that it is possible to map the real 
system ${\cal S}$ onto a fictitious system of harmonic oscillators
${\cal S}_{osc}$\,, 
\be
{\cal S} \,\, \to \,\, {\cal S}_{osc}\,,
\ee
in such a manner that $\chi_{_A}^{''} (\omega)$ and 
$\langle \hat{A}^2(\omega)\rangle$ of the real system, i.e.\,the 
imaginary part of the (Fourier transformed) generalized suscettivity
and the mean square of the (Fourier transformed) interaction operator 
$\hat A (t)$, are equivalently obtained by computing the 
corresponding quantities of the fictitious one.

It is worth to point out that the key ingredient to construct such 
a mapping is the hypothesis that linear response theory is applicable, 
which is the main hypothesis under which the FDT is established. 
When linear response theory is not valid, Eq.(\ref{resp2}) cannot 
be derived. As a consequence, we do not arrive to 
Eqs.\,(\ref{chi2}) and (\ref{o2}) 
which are crucial to build up our mapping.

Moreover, by considering the ``equivalent'' harmonic oscillators system  
${\cal S}_{osc}$ rather than the real one, we are somehow allowed to 
regard the BE distribution factor 
${\rm coth}\left( \frac{\beta\hbar\omega}{2}\right)$ 
of the FDT in Eq.\,(\ref{fdt1}) as originating from the individual
contributions ${\rm coth}\left( \frac{\beta\hbar\omega_{ji}}{2}\right)$ 
of each of the oscillators of the fictitious system (see above,
Eqs.\,(\ref{osc5}), (\ref{oscc5}) and (\ref{chi3})). 
In this sense, such a mapping allows for an oscillator 
interpretation of the BE term in the FDT.

At the same time, however, our result shows that this BE factor does 
not describe the physics of the system, i.e.\,\,it does not encode any 
real, physical, harmonic oscillator structure of the system (see also 
the considerations below). 
In this respect, we have to stress that what we have implemented  
is not a canonical transformation, i.e. it is not a transformation which 
allows to describe the system in terms of new degrees of freedom (such as normal modes), 
but just a formal mapping, a mathematical construct, which can be 
established, we repeat ourselves, only within the framework of linear 
response theory.

In our opinion, our finding provides a definite answer to the often 
raised questions of the ``physical meaning'' or ``physical origin'' 
of the BE term in the FDT or, stated differently, to the question of 
whether this BE distribution factor possibly describes the physical 
nature of the system or not\,\cite{taylor}. 

In fact, from the derivation of the FDT, we know 
that the BE factor derives from a peculiar combination of Boltzmann 
factors (see\,\cite{kubo} and Eq.\,(\ref{fddt}) above). At the same 
time, we have shown that, regardless the bosonic or fermionic 
nature of the (real) system ${\cal S}$, it is always possible to
establish a mapping which relates ${\cal S}$ to a system of harmonic 
oscillators ${\cal S}_{osc}$ so that this BE factor can be regarded 
as ``originating'' from the individual oscillators of the 
``equivalent'' system ${\cal S}_{osc}$. 
Therefore, it is not the physical nature 
of the system which is encoded in this BE term but rather a fundamental 
quantum property of any bosonic and/or fermionic system: 
{\it whenever 
linear response theory is applicable, any generic system is, at least 
with respect to the FDT, equivalent (in the sense defined above) to 
a system of quantum harmonic oscillators}.

Let us summarize now our results. We have found that, when linear 
response theory is valid, for any generic bosonic 
and/or fermionic physical system ${\cal S}$ whose interaction with an 
external field $f(t)$ is given by a term of the kind
$\hat V = - f(t) \hat A$, where $\hat A$ is an observable of the system, 
it is always possible to find an ``equivalent'' system of harmonic 
oscillators such that $\chi_{_A}^{''} (\omega)$ and 
$\langle \hat{A}^2(\omega)\rangle$ of the real system can be obtained 
by computing the corresponding quantities of the fictitious one. The 
operator that represents the physical observable $\hat A$ is such a 
mapping is the one-particle operator $\hat A_{osc}$ of 
Eq.\,(\ref{onepart}) with the $\alpha_{ji}$ and the $\omega_{ji}$ given 
by Eqs.\,(\ref{alfa}) and (\ref{omega}). 
In passing (i.e.\,in order to establish such a result), we have shown 
that for any system of 
harmonic oscillators interacting with an external field through an 
interaction term of the kind given in Eqs.\,(\ref{armint}) and 
(\ref{onepart}), the BE 
${\rm coth}\left( \frac{\beta\hbar\omega}{2}\right)$ factor originates 
from the contribution of each of the individual harmonic oscillators of 
the system. As the physical operator $\hat A$ in the mapping is 
represented by the operator $\hat A_{osc}$ of Eq.\,(\ref{onepart}), we 
concluded that the BE distribution factor which appears in the FDT 
Eq.\,(\ref{fdt1}) can be regraded as originating from the individual 
harmonic oscillators of the equivalent system. Our results clearly 
indicate that it is only in this sense that this BE factor can be 
interpreted in terms of harmonic oscillators and that no other physical 
meaning can be superimposed to it. 

We believe that our mapping has a broader range of applicability 
than the worked case of the FDT discussed in this letter. Work is in 
progress in this direction.  
\vskip 30pt
\nin
{\large{\bf Acknowledgments}}
\vskip 6pt
We would like to thank Luigi Amico, Marcello Baldo, Pino Falci and 
Dario Zappal\`a for many useful discussions.

\end{document}